\documentclass{iaus}
\usepackage{graphicx}

\title[Binaries and pulsating B and Be stars] 
{Variability of B and Be stars in the LMC/SMC: binaries and pulsations.}

\author[C. Martayan, P. Diago, J. Guti\'errez-Soto, et al.]   
{Christophe Martayan$^{1,2}$
 \and Pascual Diago$^3$
 \and Juan Guti\'errez-Soto$^{2,3}$
 \and Juan Fabregat$^3$
 \and Anne-Marie Hubert$^2$
 \and Mich\`ele Floquet$^2$
 \and Coralie Neiner$^2$
 \and Malek Mekkas$^2$  
 } 

\affiliation{$^1$Royal Observatory of Belgium, 3 avenue circulaire \\ 1180 Brussels, Belgium 
 \\ email: {\tt martayan@oma.be} \\[\affilskip]
$^2$GEPI, Observatoire de Paris, CNRS, Universit\'e Paris Diderot; 
5 place Jules Janssen \\ 92195 Meudon Cedex, France \\[\affilskip]
$^3$Observatorio Astron\'omico de Valencia, edifici Instituts d'investigaci\'o, 
Poligon la Coma, \\ 46980 Paterna Valencia, Spain
}

\pubyear{2008}
\volume{256}  
\pagerange{119--126}
\jname{The Magellanic System: Stars, Gas, and Galaxies}
\editors{Jacco Th. van Loon \& Joana M. Oliveira, eds.}
\begin{document}

\maketitle

\begin{abstract}
To study the variability of the 523 B and Be stars observed in the Magellanic clouds with the VLT-FLAMES, 
we cross-matched the stars of our sample with the photometric database MACHO, which provides for each star 
an 8 years lightcurve. 
We searched for long, medium, and short-term periodicity and found the eclipsing binaries in our sample.
For these stars, combining, spectroscopy and photometry, we were able to provide information on several systems 
of stars (systemic velocities, ratios of masses, etc). We also present the ratios of B-binaries to B-non binaries 
in the LMC/SMC in comparison with the MW. Note that this ratio is also an important issue to understand the mechanism 
of star-formation at low metallicity. 
We also found the first multiperiodic B and Be stars in the SMC, in particular the first SMC Beta Cep and SPB, while, 
according to the models, pulsations were not foreseen in low metallicity environments, i.e. typically in the SMC.
Our results show that the instability strips are shifted towards higher temperatures in comparison with the Milky Way' 
strips of pulsating B-type stars. By the fact that we found more pulsating Be stars than pulsating B stars in the SMC, 
it seems that the fast rotation favours the presence of pulsations. However, the ratio of pulsating B-type stars 
to "non"-pulsating B-type stars at low metallicity is lower than at high metallicity.

\end{abstract}

\firstsection 
\section{Introduction, observations}

To investigate the B-type star populations in the Magellanic Clouds, 
VLT-FLAMES observations of a large sample of B and Be stars were obtained in the
LMC-NGC2004 (176 stars) and SMC-NGC330 (344 stars) regions and their surroundings. We used a medium resolution
R=6400 in the blue wavelength range around H$\gamma$ (LR02) and R=8600 in the red one around H$\alpha$ (LR06). 
Among this sample 178 Be stars were observed.  We cross-correlated the coordinates of observed stars with the MACHO photometric database. We obtained the
lightcurves for the main part of stars of our sample in order to search their long-, medium-, and short-term
variability/periodicity due to potential binarity or pulsations. The time span of MACHO lightcurves is large enough to provide
a very high frequency resolution in the spectral analysis and allows us to distinguish between very close frequencies.

\section{Binaries}

\subsection{Spectroscopic and eclipsing binaries in the Magellanic Clouds}

Blue spectra of B and Be stars in the LMC were obtained 4 days before red ones and it was therefore possible to study the
variation of the radial velocities and to detect spectroscopic binaries. Despite the small number of spectra for each
object it was possible to estimate the mass ratio and the systemic velocity for some binaries (see Martayan et al. 2006a). By
cross correlation with the MACHO and OGLE databases we detected 19 photometric binaries. Among them 14 are new photometric
binaries (5 in the LMC, 9 in the SMC). Among the 19 photometric binaries, 6 are also spectroscopic binaries (5 in the LMC, 1
in the SMC) and 2 others are Be stars (2 in the SMC). For the eclipsing binaries, the orbital period was determined with an
accuracy better than 0.001d.

\subsection{Details about LMC and SMC binaries}

Thanks to observations with the VLT-GIRAFFE spectrograph, we found 23 new binary systems among the B stars of our sample in the LMC
(see Martayan et al 2006a). The orbital periods were found thanks to the MACHO data. 
In the SMC, contrary to the LMC, spectra in the blue and red wavelengths were obtained the same night
and we cannot study the variation of radial velocity. We found 13 eclipsing binaries. Their periods range from
0.664d to 455d with most common values around 2d. All the periods and details are provided in Martayan et al. (2007b). Finally
we found ~5\% of binaries among B-type stars in the SMC, ~12\% in the LMC, while in the MW, this ratio is 30\% (Porter \& Rivinius 2003).
However, the detectability of binaries in the LMC/SMC is limited and then the ratios given here are certainly lower limits.
Note that McSwain \& Gies (2005) suggested that 75\% of MW Be stars can be binaries. The identified binaries from our sample with
more complete observations of the radial velocity variation should be used for testing/constraining the input parameters of
stellar evolution models in low metallicity environments, in particular the stellar radii.

\section{Pulsating B and Be stars}

\subsection{Pulsating B and Be stars in the LMC/SMC (low metallicity)}

A significant fraction of main-sequence B type stars are variable. The B main sequence is populated by 2
classes of pulsators: the Beta Cephei stars (Stankov \& Handler 2005) and the Slowly Pulsating B stars (de Cat 2002). The
pulsations are due to the $\kappa$-mechanism acting in the partial ionisation zones of the iron-group elements. Be stars are
non-supergiant B stars whose spectrum has displayed at least once emission lines mainly in the Balmer series. Emission lines
come from a circumstellar disk created by episodic matter ejections from the central star. Be stars are also known as fast
rotator stars.  In the MW, they display short-term variations like Beta Cep or SPB stars. The $\kappa$-mechanism 
depends on the abundance of iron-group elements, and hence the respective instability strips have a great dependence 
on the metallicity of the stellar environment. Pamyatnykh (1999) showed that the Beta Cep and SPB instability strips disappear 
at Z$<$0.01 and Z$<$0.006 respectively. The LMC metallicity is Z=0.007 and the SMC one is Z=0.002 (see Maeder et al. 1999, and references therein).
Therefore it is expected to find a very low occurrence of Beta Cep and SPB pulsators in the LMC and no pulsator type in the
SMC. Previous spectroscopic study of Baade et al. (2002) failed to find variability in two Be stars, while Balona (1992) found
monoperiodic variable B-type stars in low metallicity environment. 
We downloaded MACHO lightcurves for LMC and SMC B and Be stars from the samples of Martayan et al. (2006a, 2007b).
We searched then for pulsations with these lightcurves.
The fundamental parameters based on
VLT-FLAMES spectra provided by these authors and corrected from the fast rotation effects for Be stars (see Martayan et al.
2006b, 2007a) are used. We found several B and Be stars with short-period
variability. Many of the short-period variables were found multiperiodic and some of them show a beating due to
close frequencies. We recall that this result was not expected at this low metallicity by
the theory. We propose an observational instability strip in the
SMC, which is shifted towards higher temperatures than in the MW. We also propose the hottest pulsating star in our sample as a
Beta Cep variable. The results for the MW pulsating stars come from Guti\'errez-Soto et al. (2007). The results presented here
are detailed in Diago et al. (2008).

\subsection{Ratios of pulsating B and Be stars vs. the metallicity and rotational velocities}
 
\begin{table}[h!]
  \begin{center}
  \caption{}
  \label{tab1}
  \begin{tabular}{|c|c|c|c|}\hline 
 & {\bf MW} & {\bf LMC} & {\bf SMC} \\ 
\hline
Metallicity Z & 0.020 & 0.007 & 0.002 \\
Stellar radii & decrease & with & Z \\
Linear rotational velocities & increase & with & Z \\
Critical rotational velocities & increase & with & Z \\
$\Omega$/$\Omega_{c}$ (B stars) \% & 40 & 37 & 58 \\
Pulsating B stars \% & 16 & 7 & 5 \\
$\Omega$/$\Omega_{c}$ (Be stars) \% & 80-88 & 85 & 95 \\
Pulsating Be stars \% & 74 & 15 & 25 \\
\hline
  \end{tabular}
 \end{center}
\vspace{1mm}
\end{table}

In Table~\ref{tab1}, for each galaxy, we provide the metallicity of the environment, the rates of pulsating B or Be stars, and also
their ratios of angular velocities to breakup velocities ($\Omega$/$\Omega_{c}$), which come from 
Martayan et al. (2007a, LMC and SMC), for the MW: Fr\'emat et al. (2005), Porter (1996), Chauville et al. (2001), and Stepien (2002).
When the metallicity decreases the radii of the stars decrease (see Maeder \& Meynet 2001). 
At the opposite, the linear rotational velocities of stars increase (due to
lower angular momentum loss due to lower mass-loss at low metallicity) and due to the combination of this increase with the decrease
of the radii, the critical rotational velocities also increase. The evolution of the $\Omega$/$\Omega_{c}$ is due to the combination
of these different effects.
Finally, we note: 
\begin{enumerate} 
\item a decrease of the rates of pulsating B stars with decreasing metallicity as expected by the theory.   
\item More pulsating Be stars than pulsating B stars at similar metallicity, which seems to indicate that the fast 
rotation in Be stars plays a role in the appearance of pulsations.
\item A decrease of the rates of pulsating Be stars between the MW and the LMC, which corresponds to a metallicity effect.
\item An increase of the rates of pulsating Be stars between the LMC and the SMC while the metallicity decreases. 
This may be explained by the fact that SMC Be stars rotate faster ($\Omega$/$\Omega_{c}$$\sim$95\%) 
than their counterparts in the LMC/MW ($\Omega$/$\Omega_{c}$$\sim$85\%). 
\item There is a similar trend in $\Omega$/$\Omega_{c}$ for B stars, but their ratios are always lower than the minimal 
ratio from which the fast rotation effects ($\Omega$/$\Omega_{c}$$\sim$70\%) on the stars are not negligeable (see for example
Fr\'emat et al. 2005). 
It explains why there is an effect of the fast rotation between the LMC and SMC for Be stars and not for B stars.
\end{enumerate} 

\section{Conclusion}

Our results provide new informations and constraints for the models about the pulsation theory at low metallicity. They seem to
indicate an enhancement of the non-radial pulsations due to the high rotation rates. As a consequence, due to the rotational
mixing, are the fast rotators (Be stars) enriched in metals? To confirm and enlarge our results, we need first to increase the
samples by using WFI-spectroscopic samples of Be stars in the LMC/SMC, second to determine the chemical abundances of the stars by using
high resolution spectra from VLT: UVES, X-Shooter, third to test the decrease of the stellar radii by using observations of binaries
at differetn metallicities, fourth to analyse the COROT lightcurves of MW B and Be stars, which could help
to understand the mechanisms at the origin of pulsations and the Be phenomenon.

\begin{acknowledgements}
C.M. acknowledges funding from the ESA/Belgian Federal Science Policy in the 
framework of the PRODEX program (C90290).
C.M. thanks the IAUS SOC/LOC for the IAUS grant and  the FNRS for the travel grant.
\end{acknowledgements}

\end{document}